\documentclass[aps,prl,twocolumn,showpacs]{revtex4}

\usepackage{amsmath}
\usepackage{amssymb}
\usepackage{graphicx}
\usepackage{dcolumn}
\usepackage{bm}

\newcommand{\la}{\lambda}

\newcommand{\prt}{\partial}

\begin{document}

\title{
Kinetic equation for a dense soliton gas}
\author{G.A. El$^{1}$}
\email{G.El@lboro.ac.uk}
\author{A.M. Kamchatnov$^{2}$}
\email{kamch@isan.troitsk.ru}

\affiliation{ $^1$ Department of Mathematical Sciences, Loughborough
University,
Loughborough LE11 3TU, UK \\
$^2$Institute of Spectroscopy, Russian Academy of Sciences, Troitsk,
Moscow Region, 142190, Russia }

\date{\today}

\begin{abstract}
We propose a general method to derive kinetic equations for dense
soliton gases in physical systems described by integrable
nonlinear wave equations. The kinetic equation describes evolution
of the spectral distribution function of solitons due to
soliton-soliton collisions. Owing to complete integrability of the
soliton equations, only pairwise soliton interactions contribute
to the solution and the evolution reduces to a transport of the
eigenvalues of the associated spectral problem with the
corresponding soliton velocities modified by the collisions. The
proposed general procedure of the derivation of the kinetic
equation is illustrated by the examples of the Korteweg -- de
Vries (KdV) and nonlinear Schr\"odinger (NLS) equations. As a
simple physical example we construct an explicit solution  for the
case of interaction of two cold NLS soliton gases.
\end{abstract}

\pacs{05.45.Yv}

\maketitle

The concept of  soliton plays a fundamental role in nonlinear
physics due to its main property---preservation of its parameters
during interactions with other solitons in the case of the
physically and mathematically important class of so-called
integrable equations. Such a particle-like behavior has led to a
huge number of investigations of soliton dynamics in various
physical systems (see, e.g. \cite{scott}) as well as to thorough
study of mathematical properties of integrable (or soliton)
equations (see, e.g. \cite{newell}) initiated in the pioneering
paper \cite{ggkm} where the famous inverse scattering transform
(IST) method has been formulated. In this method, each soliton is
parameterized by an eigenvalue $\la$ of the linear spectral
problem associated with the nonlinear wave equation under
consideration. For example, the  KdV equation
\begin{equation}\label{1-1}
    u_t+6uu_x+u_{xxx}=0
\end{equation}
is associated with the linear Schr\"odinger equation
\begin{equation}\label{1-2}
    \psi_{xx}+u(x,t)\psi= -\la^2\psi,
\end{equation}
so that evolution of the potential $u(x,t)$ according to
(\ref{1-1}) does not change the spectrum $\la$, and, whence, the
properties of solitons do not change either. The IST method gives
a full explanation of finite-number solitons dynamics and provides
the basis for the description of many nonlinear physics phenomena.

We encounter a different physical situation when we have to deal
with a dense lattice of solitons. When the solitons in the lattice
are correlated they may form a modulated nonlinear periodic wave.
The spectral problem (\ref{1-2}) for a general periodic in $x$
``potential'' $u(x,t)$ leads to the Bloch band structure of the
spectrum $\la$. The periodic  ``soliton lattices" are
distinguished by a finite number of gaps in the spectrum (see
\cite{nov}) and the edges $\la_i$ of the spectral gaps become
convenient parameters in terms of which the major physical
characteristics of the wave such as wavelength, frequency,
amplitude  etc are expressed. In a weakly modulated wave the
parameters $\la_i$ become slow functions of space and time
coordinates, and their evolution is governed by the Whitham
equations \cite{whitham}. Methods of derivation and integration of
the Whitham equations are now well developed and this theory
provides an adequate description of such important phenomenon as
formation of dispersive shocks (or undular bores) in various
physical systems from water surface to space plasma and
Bose-Einstein condensate.

Yet another principally different class of problems arises when
solitons form a disordered finite-density ensemble (a soliton gas)
rather than well-ordered modulated soliton lattice. The relevant
physical conditions can be realized for instance when large number
of solitons are generated either by random large-scale initial
distributions \cite{osb} or by a stochastic external forcing (such
as irregular topography in the internal water wave dynamics), or
when they are injected into a ring resonator \cite{mitsch}. Here
it is necessary to introduce an appropriate kinetic description of
the soliton gas. We introduce a distribution function $f(x,t;\la)$
as the number of solitons with the spectral parameter $\la$ in the
interval $(\la,\la+d\la)$ and in the space interval $(x,x+dx)$ at
the moment $t$. Now, if the soliton dynamics is governed by an
exactly integrable equation  we arrive at the problem of
describing the isospectral evolution of the distribution function
$f(x,t;\la)$ with time. Since the spectrum $\la$ is preserved, the
evolution of $f(x,t;\la)$ must be governed by the conservation
equation
\begin{equation}\label{2-1}
    f_t+(sf)_x=0 \, ,
\end{equation}
which means that  the eigenvalues $\la$ are transferred along the
$x$-axis with some mean velocity $s(x,t;\la)$ depending on the
distribution function $f$ but there is no exchange of $\la$'s
between different $\la$-intervals. Hence, the problem is reduced
to finding the velocity $s$ of a soliton gas as a function of
$\la$, $x$ and $t$. This problem was posed by Zakharov
\cite{zakh71} as early as in 1971 and he solved it for the case of
rarefied gas of KdV solitons. Only recently it was generalized in
Ref.~\cite{el03} to the case of a dense gas of KdV solitons where
it was found that velocity $s(x,t;\la)$ is determined from the
integral equation
\begin{equation}\label{ints0}
s(\la)=4\la^2+\frac{1}{\la}\int \limits^{\infty}_0 \ln
\left|\frac{\la + \mu}{\la-\mu}\right|f(\mu)[s(\la)-s(\mu)]d\mu,
\end{equation}
so that Eqs.~(\ref{2-1}) and (\ref{ints0}) give a closed
self-consistent kinetic description of a soliton gas of an
arbitrary density. In the limit of rarefied gas, i.e. for
$
    \int f(\mu)d\mu\ll \la_0,
$
where $\la_0$ is a characteristic value of the spectral
parameter, the second term in Eq.~(\ref{ints0}) becomes a small
correction to the speeds $4\la^2$ of non-interacting solitons, and
the substitution of $s(\eta)\cong 4\eta^2,$ $(\eta=\la,\mu)$,
reproduces the Zakharov kinetic equation \cite{zakh71}.

Although a mathematically rigorous derivation of Eq.~(\ref{ints0})
given in Ref.~\cite{el03} (which is based on a certain singular
limit of the Whitham equations) is quite technical, the final
result is physically very natural and suggestive. Indeed,
Eq.~(\ref{ints0}) implies that only two-soliton collisions have to
be taken into account  (which agrees with the properties of
multi-soliton solutions  of the KdV equation); thus in a collision
of a $\la$-soliton (i.e. with the spectral parameter $\la$) with a
$\mu$-soliton, the coordinate of the $\la$-soliton
is shifted by the distance
$$
     \frac{1}{\la}\ln
\left|\frac{\la + \mu} {\la-\mu}\right|\quad\text{for}\quad\la>\mu,
$$
(and there is a  similar expression for $\la<\mu$), and number of
collisions per second  is proportional to the relative mean
velocity $[s(\mu)-s(\la)]$ of these two types of solitons
multiplied by the density of $\mu$-solitons. Then, after
integration over the distribution function $f(\mu)$ of
$\mu$-solitons, we arrive at equation (\ref{ints0}) for the speed
of $\la$-solitons modified by their collisions with the other
$\mu$-solitons. It is supposed that the number of the soliton
collisions over large distance is large enough and, hence, the
mean velocity $s$ is a well-defined variable. As a matter of fact,
the typical $x,t$-scales in the kinetic equation are much larger
than in the original equation (\ref{1-1}). Thus, one can see that
the above simple reasoning provides an independent derivation of
Eq.~(\ref{ints0}) for the KdV equation case and, obviously, it can
be directly applied to other integrable equations. In this Letter
we use this method to derive the kinetic equation for a
finite-density gas of bright NLS solitons and infer some its
consequences.

As is known, each soliton solution of the focusing NLS equation
\begin{equation}\label{eq1}
    iu_t+u_{xx}+2|u|^2u=0
\end{equation}
is characterized by a complex eigenvalue
\begin{equation}\label{eq2}
    \la=\alpha+i\gamma,\quad -\infty<\alpha<\infty,\quad
    0<\gamma<\infty,
\end{equation}
of the Zakharov-Shabat spectral problem and is given by (see, e.g. \cite{nov})
\begin{equation}\label{eq3}
    u(x,t)=2i\gamma\frac{\exp[-2i\alpha
x-4i(\alpha^2-\gamma^2)t-i\phi_0]}
    {\cosh[2\gamma(x+4\alpha t-x_0)]},
\end{equation}
that is $\gamma=\Im(\la)$ determines the amplitude of the soliton
and $\alpha=\Re(\la)$ its velocity $v=-4\alpha$. The multi-soliton
solution shows that interaction of solitons reduces to only
two-soliton elastic collisions and effects of multi-soliton
collisions are absent. In accordance with the above argumentation,
we consider a gas of solitons characterized by a continuous
distribution function of eigenvalues $\la$. In other words, $
f(x,t;\alpha,\gamma)d\alpha d\gamma dx$ is the number of
eigenvalues in the element $d\alpha d\gamma$ of the complex plane
$\la$ in the space interval $dx$ much greater than both typical
soliton width $\sim1/\gamma$ and average distance between
solitons. This distribution function evolves due to motion of
solitons and as a consequence of preservation of the spectrum it
satisfies again the continuity equation (\ref{2-1}) which means
conservation of the density of eigenvalues in a given element of
$\la$-plane. In this equation $s$ denotes mean velocity of
$\la$-solitons which should be evaluated with the account of
soliton collisions.  Without interaction of solitons it would be
equal to $s(\alpha,\gamma)=-4\alpha$. However, collisions of
$\la$-soliton $(\la=\alpha+i\gamma)$ with other $\mu$-solitons
$(\mu=\xi+i\eta)$ modify it in the following way. Each collision
of faster $\la$-soliton with slower $\mu$-soliton shifts the
$\la$-soliton forward to the distance (see, e.g. \cite{nov})
$$
\frac1{2\gamma}\ln\left|\frac{\la-\overline{\mu}}{\la-\mu}\right|^2
\quad \mathrm{for} \quad s(\alpha,\gamma)>s(\xi,\eta),
$$
and the number of such collisions in the time interval $dt$ is
equal to the product of the density $ f(\xi,\eta)d\xi d\eta $ and
the distance overcame by faster $\la$-solitons compared with
slower $\mu$-solitons, $ [s(\alpha,\gamma)-s(\xi,\eta)]dt$. Thus,
such collisions increase a path covered by $\la$-soliton compared
with $-4\alpha dt$. In a similar way, one can calculate the
negative shift due to collisions with faster $\mu$-solitons. The
total shift is obtained by integration over $d\xi d\eta$. Equating
paths $s(\alpha,\gamma)dt$ and $-4\alpha dt +\mathrm{``total\,\,
shift"}$, we obtain an integral equation for the self-consistent
definition of the soliton gas velocities,
\begin{equation}\label{eq9}
\begin{split}
    s(\alpha,\gamma)=-4\alpha
    +\frac1{2\gamma}\int_{-\infty}^\infty\int_0^\infty
    \ln\left|\frac{\la-\overline{\mu}}{\la-\mu}\right|^2f(\xi,\eta)\\
    \times[s(\alpha,\gamma)-s(\xi,\eta)]d\xi d\eta.
    \end{split}
\end{equation}
Obviously, this derivation of Eq.~(\ref{eq9}) is correct as long
as the resulting velocity $s(\alpha,\gamma)$ is finite. Under this
reservation, equations (\ref{2-1}) and (\ref{eq9}) provide the
basis for a consideration of kinetic behavior of dense gas of NLS
solitons.

As a simple application of the kinetic equations, let us consider
evolution of two beams of solitons, when the spectral
distribution gas consists of two monochromatic (``cold'') parts:
\begin{equation}\label{eq10}
\begin{split}
    f(x,t;\xi,\eta)&=\rho_1(x,t)\delta(-\xi-\alpha,\eta-\gamma)\\
    &+    \rho_2(x,t)\delta(\xi-\alpha,\eta-\gamma),
    \end{split}
\end{equation}
where $\rho_1(x,t)$ corresponds to fast (or moving to the right in
the reference system associated with the group velocity of the
carrier wave) solitons ($\la_1=-\alpha+i\gamma$) and $\rho_2(x,t)$
to slow (or moving to the left) solitons ($\la_2=\alpha+i\gamma$).
All solitons have the same amplitudes. Of course, the idealized
delta-functional approximation  for the distribution function in
the kinetic equation means that the exact soliton eigenvalues in
the original spectral problem for the  NLS equation are
distributed for $i$-th component within a narrow vicinity of the
dominant value $\lambda=\lambda_i$ and are actually different,
which precludes formation of bound states. The soliton positions
in such a ``monochromatic'' gas are statistically independent
which leads to Poisson distribution for the number of solitons in
a unit space interval with the mean density $\rho_i$ \cite{ekmv}.
It is also clear that one can neglect interactions between the
solitons belonging to the same beam  compared with the cross-beam
interactions.

Substituting (\ref{eq10}) into (\ref{2-1}), (\ref{eq9}) we obtain
a ``two-beam'' reduction of the kinetic equation,
\begin{equation}\label{eq14}
    \frac{\prt \rho_1}{\prt t}+\frac{\prt (s_1\rho_1)}{\prt x}=0,\quad
    \frac{\prt \rho_2}{\prt t}+\frac{\prt (s_2\rho_2)}{\prt x}=0,
\end{equation}
where the velocities of the beams $s_{1}=s(-\alpha, \gamma)$ and
$s_2=s(\alpha, \gamma)$ are determined from the equations
\begin{equation}\label{eq11}
    s_1=4\alpha+\kappa \rho_2\cdot (s_1-s_2),\ \
    s_2=-4\alpha+\kappa \rho_1 \cdot (s_2-s_1),
\end{equation}
and the interaction parameter
\begin{equation}\label{eq12}
    \kappa=\frac1{2\gamma}\ln\left(1+\frac{\gamma^2}{\alpha^2}\right)
\end{equation}
is always positive. Resolving (\ref{eq11}) we get expressions for
velocities in terms of $\rho_{1,2}(x,t)$:
\begin{equation}\label{eq13}
s_1=4\alpha\frac{1-\kappa(\rho_1-\rho_2)}{1-\kappa(\rho_1+\rho_2)},\
\
s_2=-4\alpha\frac{1+\kappa(\rho_1-\rho_2)}{1-\kappa(\rho_1+\rho_2)}.
\end{equation}
 According to the formulated above
condition of applicability of the kinetic description, the
densities must satisfy the inequality
\begin{equation}\label{4-1}
    \kappa(\rho_1+\rho_2)<1.
\end{equation}
Substitution of the inverse expressions for $\rho_{1,2}$,
\begin{equation}\label{eq16}
\rho_1=\frac{s_2+4\alpha}{\kappa (s_2-s_1)}\, , \qquad
\rho_2=\frac{s_1-4\alpha}{\kappa(s_1-s_2)}
\end{equation}
into the system (\ref{eq14}) reduces it to the
form
\begin{equation}\label{eq15}
\frac{\partial s_1}{\partial t}+s_2\frac{\partial s_1}{\partial
x}=0 \, , \qquad \frac{\partial s_2}{\partial t}+s_1\frac{\partial
s_2}{\partial x}=0,
\end{equation}
which is known as the Riemann invariant form of a Chaplygin gas
equations which, besides well-known original application to
compressible gas dynamics, has recently found a number of other
applications. If solution of the system (\ref{eq15}) is known,
then the densities $\rho_1, \rho_2$ are determined in terms of
$s_1, s_2$ by the formulas (\ref{eq16}).

Although the system (\ref{eq15}) admits the general solution (see,
e.g. \cite{fer}), we shall confine ourselves to a physically
natural problem of the collision (mixing) of slow and fast soliton
gases when two gases are separated  at the initial moment so that
\begin{equation}\label{icf1}
\rho_1(x,0)= \rho_{10}(x) ,  \quad \rho_2(x,0)= 0 \quad \qquad
\hbox{for} \
\ x<0\, ,
\end{equation}
\begin{equation}\label{icf2}
\rho_1(x,0)= 0 ,  \qquad  \rho_2(x,0)= \rho_{20}(x) \ \quad \hbox{for}
\
\ x>0\,
\end{equation}
where $\rho_{10}(x)>0$ and $\rho_{20}(x)>0$ are given functions.
Corresponding initial data for the
equivalent system (\ref{eq15}) follow from (\ref{eq13}):
\begin{equation}
s_1(x,0)= 4\alpha,\ \  \ s_2(x,0)=
-4\alpha\frac{1+\kappa\rho_{10}(x)}{1-\kappa\rho_{10}(x)},
\label{5-1}
\end{equation}
for $x<0$, and
\begin{equation}\label{5-2}
s_1(x,0)=
4\alpha\frac{1+\kappa\rho_{20}(x)}{1-\kappa\rho_{20}(x)},\ \ \
s_2(x,0)= -4\alpha,
\end{equation}
for $x>0$. To further simplify the problem, we consider the case
when both gases are homogeneous, i.e.,  $\rho_{10}(x)=\rho_{10}$,
$\rho_{20}(x)=\rho_{20}$, where $\rho_{10}$ and $\rho_{20}$ are
constant. Since in this case neither the initial conditions, nor
the evolution equations (\ref{eq15}) contain any parameter with
dimension of length, the solution must be self-similar and depend
on the variable $x/t$ alone. However, as can be easily seen, the
system (\ref{eq15}) does not possess non-constant similarity
solutions. On the other hand, the solution consisting of two
constants $s_1\, , s_2$ cannot satisfy the discontinuous initial
conditions (\ref{5-1}), (\ref{5-2}). As in  shock wave theory
\cite{whitham}, this can be remedied by introducing admissible
discontinuities in the solution. The  discontinuous ``weak''
solutions are allowed here owing to the presence of the
conservation laws (\ref{eq14}), (\ref{eq13}). As a result, the
sought solution has the form of three constant states
$\{f_1,f_2\}: \{\rho_{10},0\}, \{\rho_{1c}, \rho_{2c} \},
 \{ 0,\rho_{20}\} $ separated by two jump discontinuities.

We denote the velocities of the discontinuities as $c^{\pm}\, ,
\quad c^+>c^-$. Then the solution is:

(i) $x<c^-t$:

\begin{equation}\label{sol-}
 \rho_1\equiv \rho_1^-=\rho_{10}\, ,
 \quad \rho_2 \equiv \rho_2^- =0\, ,
\end{equation}
which implies by (\ref{eq13}),
\begin{equation}\label{sols-}
 s_1 \equiv s_1^-=4\alpha\, , \quad s_2 \equiv
s_2^-=-4\alpha\frac{1+\kappa\rho_{10}}{1-\kappa\rho_{10}}
\end{equation}
 (ii) $x>c^+t$:
\begin{equation}\label{sol+}
 \rho_1\equiv \rho_1^+=0 , \quad \rho_2 \equiv \rho_2^+
=\rho_{20} ,
\end{equation}
which implies
\begin{equation}\label{sols+}
s_1 \equiv s_1^+=4\alpha\frac{1+\kappa\rho_{20}}{1-\kappa\rho_{20}}\, ,
\quad s_2 \equiv s_2^+=-4\alpha  .
\end{equation}
The values $s_2^-$ and $s_1^+$ can be viewed as velocities of
trial solitons of one component moving through the homogeneous gas
of solitons of another component. One can see that there are
critical values for the densities  $\rho_{10} =\kappa^{-1}$ and
$\rho_{20} =\kappa^{-1}$ yielding the infinite speeds for trial
solitons. In accordance with the restriction described above, we
assume $\kappa \rho_{10} <1$, $\kappa\rho_{20} <1$.

(iii)  $c^-t<x<c^+t$:

Let $\rho_1 \equiv \rho_{1c} , \,  \rho_2 \equiv \rho_{2c}$ in this
interaction zone. Then the four unknown values $\rho_{1c}, \rho_{2c},
c^+, c^-$ are found from the jump conditions consistent with the
physical conservation laws (\ref{eq14}) (see, for instance,
\cite{whitham}):
\begin{equation}
\begin{array}{l}
-c^-[\rho_{1c}-\rho_1^-]+[\rho_{1c} s_{1c}-\rho_1^- s_1^-]=0  , \\
-c^-[\rho_{2c}-\rho_2^-]+[\rho_{2c} s_{2c}-\rho_2^- s_2^-]=0 ,
\end{array}
\label{j1}
\end{equation}
\begin{equation}
\begin{array}{l}
-c^+[\rho_{1c}-\rho_1^+]+[\rho_{1c} s_{1c}-\rho_1^+ s_1^+]=0 , \\
-c^+[\rho_{2c}-\rho_2^+]+[\rho_{2c} s_{2c}-\rho_2^+ s_2^+]=0  .
\end{array}
\label{j2}
\end{equation}
Here
\begin{equation}\label{s120}
s_{1c} =4\alpha\frac{1-\kappa(\rho_{1c}- \rho_{2c})}{1-\kappa
(\rho_{1c}+\rho_{2c})}, \ \ \ s_{2c} = -4\alpha\frac{1+\kappa
(\rho_{1c}-\rho_{2c})}{1-\kappa (\rho_{1c}+\rho_{2c})}.
\end{equation}
In view of (\ref{sol-}), (\ref{sol+}) it readily follows from
(\ref{j1}) and (\ref{j2}) that $c^-=s_{2c}$ and $c^+=s_{1c}$. Then
$\rho_{1c}$,
$\rho_{2c}$ are  found from the same equations with the account of
(\ref{sol-}), (\ref{sol+}) as
\begin{equation}\label{eq25}
\rho_{1c} = \frac{\rho_{10}(1-\kappa
\rho_{20})}{1-\kappa^2\rho_{10}\rho_{20}},
\quad \rho_{2c} = \frac{\rho_{20}(1-\kappa
\rho_{10})}{1-\kappa^{2}\rho_{10}\rho_{20}}  .
\end{equation}
Correspondingly, substitution of (\ref{eq25}) into (\ref{s120})
yields the velocities of expansion of the interaction
region
\begin{equation}\label{eq26}
 c^-=s_{2c}= -4\alpha
\frac{1+\kappa\rho_{10}}{1-\kappa\rho_{10}}
\quad c^+=s_{1c}= 4\alpha
\frac{1+\kappa\rho_{20}}{1-\kappa\rho_{20}},
\end{equation}
and this completes the solution. As one should expect, these
velocities coincide with velocities $s_2^-$ and $s_1^+$ of the
trial solitons (see Eqs.~(\ref{sols-}) and (\ref{sols+})). Thus,
we have found boundaries $x=c^{\pm}t$ of the interaction region
and densities $\rho_{1c},$ $\rho_{2c}$ of the two components of
soliton gas in this region. In particular, the obtained solution
in view of (\ref{4-1}) implies that the following inequality is
satisfied:
\begin{equation}\label{F}
F(\rho_{1c},
\rho_{2c})=\frac{\rho_{1c}+\rho_{2c}}{\rho_{10}+\rho_{20}}=
\frac{1-\frac{2\kappa\rho_{10}\rho_{20}}{\rho_{10}+\rho_{02}}}
{1-\kappa^2\rho_{10}\rho_{20}}
<1  .
\end{equation}
It means that the total density in the interaction (mixing) zone
is always less than the sum of densities of individual separated
components.

\begin{figure}[ht]
\centerline{\includegraphics[width=7cm,height=
4.5cm,clip]{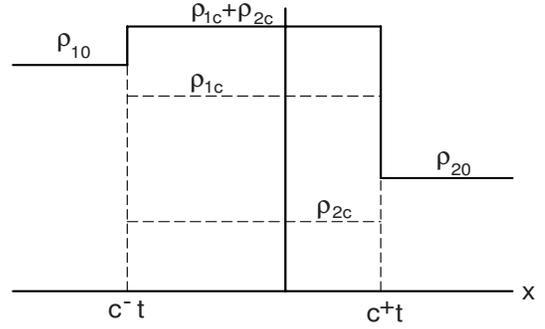}}
\caption{Dependence of solitons densities on space coordinate $x$
at some moment $t$; $\rho_{10}$ and $\rho_{20}$ are the densities of
solitons
in the beams propagating to the left and to the right, correspondingly;
$\rho_{1c}$ and $\rho_{2c}$ are the densities of the same beams in the
overlap region.}  \label{fig1}
\end{figure}

The process of the collision of two soliton gases is illustrated
in Fig.~1. Due to the interaction of solitons with each other, the
overlap region spreads out faster than it would without taking
into account the phase-shifts caused by two-soliton collisions,
and the kinetic equation allows one to give quantitative
description of this effect.

In conclusion, we have obtained the kinetic equation describing
the evolution of the spectral distribution function of a dense gas
of uncorrelated NLS solitons. The proposed procedure of the
derivation can be generalized to whole AKNS hierarchy and to other
integrable hierarchies. A process of interaction of two ``cold''
NLS soliton gases is studied in detail.

We are grateful to the Referees for a number of valuable
comments. AMK thanks RFBR (grant 05-02-17351) for financial
support.


\begin{thebibliography}{99}

\bibitem{scott} A.C. Scott, {\it Nonlinear Science: Emergence and
Dynamics
of Coherent Structures,} (Oxford University Press, Oxford, 2003).

\bibitem{newell} A.C. Newell, {\it Solitons in Mathematics and
Physics,}
(SIAM, Philadelphia, 1985).

\bibitem{ggkm} C.S. Gardner, J.M. Greene, M.D. Kruskal, and R.M. Miura,
Phys. Rev. Lett. {\bf 19,} 1095 (1967).

\bibitem{nov} S.P. Novikov, S.V. Manakov, L.P. Pitaevskii,
 and V.E. Zakharov, {\it Theory of Solitons: The Inverse
 Scattering Transform,} (Plenum, New York, 1984).

\bibitem{whitham} G.B. Whitham,
{\it Linear and Nonlinear Waves,} (Wiley--Interscience, New York,
1974).

\bibitem{osb} A.R. Osborne, Phys. Rev. Lett. {\bf 71,} 3115 (1993).

\bibitem{mitsch} F. Mitschke, I. Halama, and A. Schwache, Chaos,
Solitons, Fractals, {\bf 10,} 913 (1999).

\bibitem{zakh71} V.E. Zakharov, Zh. Eksp. Teor. Fiz. {\bf 60,} 993
(1971).

\bibitem{el03} G.A. El, Phys. Lett. A {\bf 311,} 374 (2003).

\bibitem{ekmv} G.A. El, A.L. Krylov, S.A. Molchanov, and S.
Venakides, Physica D {\bf 152,} 653  (2001).

\bibitem{fer} E.V. Ferapontov, Phys. Lett. A {\bf 158,} 112 (1991).


 \end{thebibliography}
\end{document}